\documentclass[conference]{IEEEtran}
\IEEEoverridecommandlockouts
\usepackage{amsmath}
\usepackage{amssymb}
\usepackage{amsfonts}
\usepackage{algorithm}
\usepackage{algpseudocode}
\usepackage{comment}
\usepackage{dsfont}
\usepackage{float}
\usepackage{cite}
\usepackage{graphicx}
\usepackage[left=0.625in,right=0.625in,top=0.75in,bottom=1in]{geometry}
\usepackage{epsfig}
\usepackage{subfigure}
\usepackage{psfrag}
\usepackage{xcolor}
\usepackage{url}
\allowdisplaybreaks
\usepackage[colorlinks,linkcolor=black,urlcolor=black,anchorcolor=black,citecolor=black,hyperfootnotes=true]{hyperref}

\newtheorem{remark}{Remark}

\begin{document}

\title{MUSIC Algorithm for IRS-Assisted AOA Estimation}	
\author{\IEEEauthorblockN{Qipeng Wang,
Liang Liu,
and Shuowen Zhang}
\IEEEauthorblockA{Department of Electrical and Electronic Engineering, The Hong Kong Polytechnic University \\ E-mails: qipeng.wang@connect.polyu.hk, $\{$liang-eie.liu,shuowen.zhang$\}$@polyu.edu.hk}
\thanks{This work was supported by the Research Grants Council, Hong Kong, China, under Grant 25215020, Grant 15203222, and Grant 15230022.}}

\maketitle	
	
\begin{abstract}
Based on the signals received across its antennas, a multi-antenna base station (BS) can apply the classic multiple signal classification (MUSIC) algorithm for estimating the angle of arrivals (AOAs) of its incident signals. This method can be leveraged to localize the users if their line-of-sight (LOS) paths to the BS are available. In this paper, we consider a more challenging AOA estimation setup in the intelligent reflecting surface (IRS) assisted integrated sensing and communication (ISAC) system, where LOS paths do not exist between the BS and the users, while the users' signals can be transmitted to the BS merely via their LOS paths to the IRS as well as the LOS path from the IRS to the BS. Due to the lack of the LOS paths between the BS and the users, we treat the IRS as the anchor and are interested in estimating the AOAs of the incident signals from the users to the IRS. Note that we have to achieve the above goal based on the signals received by the BS, because the passive IRS cannot process its received signals. However, the signals received across different antennas of the BS only contain AOA information of its incident signals via the LOS path from the IRS to the BS, which is not helpful for localizing the users. To tackle this challenge arising from the spatial-domain received signals, we propose an innovative approach to create temporal-domain multi-dimension received signals for estimating the AOAs of the paths from the users to the IRS. Specifically, via a proper design of the user message pattern and the IRS reflecting pattern, we manage to show that our designed temporal-domain multi-dimension signals can be surprisingly expressed as a function of the virtual steering vectors of the IRS towards the users. This amazing result implies that the classic MUSIC algorithm can be applied to our designed temporal-domain multi-dimension signals for accurately estimating the AOAs of the signals from the users to the IRS, even if these signals are received and processed by the BS, rather than the passive IRS. This new finding is verified by numerical results.
\end{abstract}

\section{Introduction}\label{Sec:intro}
Radar and wireless communication are acknowledged as the two most successful applications of the radio technology. Previously, these two systems were independently designed and implemented with limited hardware intersection. However, there is a recent trend to achieve integrated sensing and communication (ISAC) just in one common system \cite{isac_survey1,isac_survey2}. Currently, there is a consensus that ISAC will be of paramount importance to the sixth-general (6G) cellular network \cite{Tan21}.

This paper considers angle of arrival (AOA) estimation in 6G ISAC network. Previously, AOA estimation has been widely studied when the line-of-sight (LOS) paths between the users and the multi-antenna base station (BS) exist \cite{angle_survey}. Under this setup, plenty of AOA estimation algorithms have been proposed in the literature, e.g., Capon beamforming \cite{Capon}, estimation of signal parameters via rational invariance techniques (ESPRIT) \cite{ESPRIT}, compressed sensing based methods \cite{Compressed_sensing}, and multiple signal classification (MUSIC) \cite{music}. However, LOS paths may not exist between the users and the BS in practice. In these scenarios, the angular information of the users cannot be estimated via the signals received by the multi-antenna BS. Recently, the intelligent reflecting surface (IRS) has emerged as a promising solution to enhance the network throughput of 6G cellular systems \cite{irs_survey2,irs_survey}. This throughput gain is more significant when an IRS is deployed at a proper location such that it possesses both LOS paths to the user side and to the BS side. Although this property has been widely utilized for high-speed communication, we aim to emphasize that it is also beneficial for AOA estimation in sensing. Specifically, the IRS, whose location is known, can be treated as anchor, and we may estimate the AOAs of the signals via the LOS paths from the users to the IRS for localizing the users.

\begin{figure}
\centering
\includegraphics[scale=0.09]{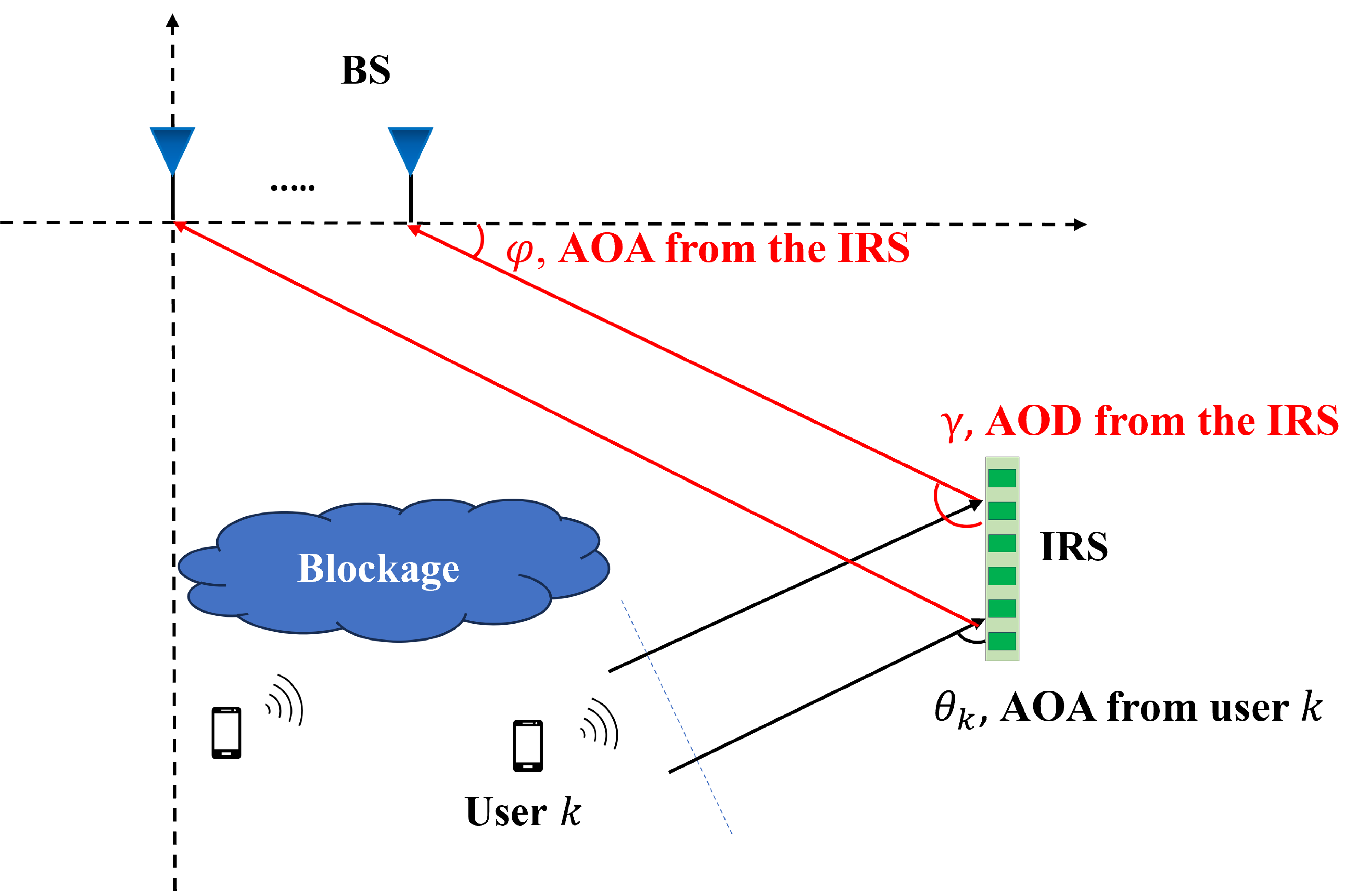}
\caption{System model for AOA estimation in IRS-assisted ISAC network: The AOAs from the users to the BS cannot be estimated because no LOS paths exist between them. Instead, we aim to estimate the AOAs from the users to the IRS, but based on the signals received by the BS.}\vspace{-10pt}
\label{fig1}
\end{figure}

To achieve the above goal, this paper considers the AOA estimation problem in an IRS-assisted ISAC system as shown in Fig. \ref{fig1}, which consists of a multi-antenna BS, an IRS, and multiple single-antenna users. We assume that the LOS paths between the users and the BS do not exist, but those between the users and the IRS and between the IRS and the BS exist. Under this setup, the users transmit uplink signals to the BS via the assistance of the IRS, while besides to decode user messages, the BS also aims to estimate the angular information of the users based on its received signals. The main challenge is as follows - we aim to estimate the AOAs of the users towards the IRS, but the signals received across the antennas at the BS on different time samples are merely useful for estimating the AOA of its incident signals, i.e., the AOA from the IRS to the BS. To tackle this challenge, we propose a novel approach to create temporal-domain multi-dimension received signals via a proper design of the user message pattern and the IRS reflecting pattern. Different from the spatial-domain received signals used in conventional AOA estimation works, our designed temporal-domain signals can be surprisingly expressed as a function of the virtual steering vectors of the IRS towards the users. Based on this property, we manage to apply the classic MUSIC algorithm on the temporal-domain signals for accurately estimating the AOAs of the signals from the users to the IRS.

We want to emphasize that there are several works considering the utilization of the IRSs as the anchors for the sensing purpose. Specifically, in \cite{IRS_ISAC1}, we proposed an algorithm that can estimate the distance between each user and the IRS accurately, even if the passive IRS cannot directly estimate such information. Moreover, AOA estimation in IRS-assisted network has been studied in \cite{aoa,irs_aoa_toa1}. However, these works focus on localizing one user, while our paper can be applied to multi-user localization. More importantly, these papers do not reveal the feasibility of utilizing MUSIC for AOA estimation in IRS-assisted ISAC network. Our main contribution over the above works is to show that we can apply the MUSIC algorithm to estimate the AOAs of the signals from the users to the IRS even if the passive IRS cannot process its received signals across different reflecting elements and we merely have the signals received at the BS.

\section{Preliminary About the MUSIC Algorithm}
Although the MUSIC algorithm is very famous, it is beneficial to list some necessary conditions under which this algorithm works for AOA estimation. These conditions will be helpful to illustrate the challenge for AOA estimation in our considered IRS-assisted ISAC system later in the paper.

Suppose there are $K$ single-antenna users and one BS equipped with $M$ antennas. Let $\theta_k$ denote the AOA from user $k$ to the BS, and $\boldsymbol{a}(\theta_k)\in \mathbb{C}^{M\times 1}$ denote the steering vector for user $k$, $\forall k$. Then, the received signal of the BS at time sample $n$ is given by
\begin{align}\label{eq:received signal MIMO}
    \boldsymbol{y}^{(n)} =  \sum_{k=1}^{K}   \beta_k\boldsymbol{a}(\theta_k) \sqrt{p_k}s_{k}^{(n)} + \boldsymbol{z}^{(n)} =\boldsymbol{A}(\boldsymbol{\theta})\boldsymbol{x}^{(n)}+\boldsymbol{z}^{(n)},
\end{align}where $p_k$ denotes the transmit power of user $k$, $s_{k}^{(n)}$ denotes the message of user $k$ transmitted at time sample $n$, $\boldsymbol{z}^{(n)} \sim \mathcal{CN}(\boldsymbol{0},\sigma^2\boldsymbol{I})$ denotes the noise at the BS at time sample $n$, $\boldsymbol{\theta}=[\theta_1,\cdots,\theta_K]^T$, $\boldsymbol{A}(\boldsymbol{\theta})\overset{\Delta}{=}[\boldsymbol{a}(\theta_1),\cdots,\boldsymbol{a}(\theta_K)]$, and $\boldsymbol{x}^{(n)}\overset{\Delta}{=}[\beta_1 \sqrt{p_1}s_{1}^{(n)},\cdots,\beta_K\sqrt{p_K}s_{K}^{(n)}]^T$. Define $\boldsymbol{Q}$ and $\boldsymbol{R}$ as the covariance matrices of $\boldsymbol{x}^{(n)}$'s and $\boldsymbol{y}^{(n)}$'s, respectively, such that $\boldsymbol{R}=\boldsymbol{A}(\boldsymbol{\theta})\boldsymbol{Q}\boldsymbol{A}^H(\boldsymbol{\theta})+\sigma^2\boldsymbol{I}$. Then, we can apply the classic MUSIC algorithm to estimate the AOAs of $K$ users, i.e., $\theta_1,\cdots,\theta_K$, under the following three necessary conditions:
\begin{align}
& M>K, \label{eqn:C1} \\
& {\rm rank}(\boldsymbol{A}(\boldsymbol{\theta})\boldsymbol{Q}\boldsymbol{A}^H(\boldsymbol{\theta}))=K, \label{eqn:C2} \\
& \boldsymbol{A}(\boldsymbol{\theta}) ~ {\rm does ~ not ~ change ~ over ~ time}. \label{eqn:C3}
\end{align}After listing these necessary conditions, we focus on the AOA estimation problem in IRS-assisted ISAC network.

\section{System Model}\label{Sec:SysMod}

We consider an IRS-assisted ISAC system as illustrated in Fig. \ref{fig1}, which consists of one BS equipped with $M$ antennas, one IRS equipped with an $I$ passive elements, and $K$ single-antenna users. In the uplink, the users transmit independent data signals to the BS via the assistance of the IRS. Then, the BS can simultaneously decode the user messages for communication and estimate the angular information of the users for localization based on its received signals. Since IRS-assisted communication has been widely studied in the literature, this paper mainly focuses on IRS-assisted sensing to achieve ISAC.

In this paper, we consider a challenging localization scenario where the LOS paths between the BS and the users are blocked such that the conventional methods for estimating the AOAs of the LOS paths from the users to the BS (e.g., MUSIC) are not applicable. Instead, because the IRS is at a fixed and known location, we aim to treat the IRS as an anchor and leverage the signals received by the BS to estimate the AOAs of the LOS paths from the users to the IRS for the localization purpose. Specifically, we assume that the BS and the users are all in the far-field region of the IRS. Then, given any $\eta\in [0,2\pi)$, the steering vector of the IRS towards $\eta$ and that of the BS towards $\eta$ are denoted as $\boldsymbol{b}(\eta)\in \mathbb{C}^{I\times 1}$ and $\boldsymbol{c}(\eta)\in \mathbb{C}^{M\times 1}$, respectively. For example, if we consider a uniform linear array (ULA) model for the IRS and the BS where the spacing between the adjacent IRS elements is $d_{\text{I}}$ meters and that between the adjacent antennas is $d_{\text{B}}$ meters, then the steering vectors of the IRS and the BS towards $\eta$ are given by
\begin{align}
    & \boldsymbol{b}^{\rm ULA}(\eta) = [1,\cdots,e^{-j\frac{2\pi (I-1) d_{\text{I}}}{\lambda}\text{cos}(\eta)}]^T, \label{eqn:ULA1} \\
    & \boldsymbol{c}^{\rm ULA}(\eta) = [1,\cdots,e^{-j\frac{2\pi (M-1) d_{\text{B}}}{\lambda}\text{cos}(\eta)}]^T, \label{eqn:ULA2}
\end{align}where $\lambda$ is the wavelength. Let $\theta_k$ and $\beta_k$ denote the AOA and the path loss factor for the LOS path from user $k$ to the IRS, $\forall k$. Then, the LOS channel from user $k$ to the IRS is given by
\begin{align}
    \boldsymbol{h}_k = \beta_k \boldsymbol{b}(\theta_k)\in \mathbb{C}^{I\times 1},\quad \forall k.
\end{align}Similarly, let $\gamma$, $\varphi$, and $\delta$ denote the angle of departure (AOD), AOA, and the path loss factor for the LOS path from the IRS to the BS. Note that $\gamma$ and $\varphi$ are known because the locations of the BS and the IRS are known. Then, the LOS channel from the IRS to the BS is given by
\begin{align}\label{eq:channel IRS BS}
    \bar{\boldsymbol{G}} = \delta\boldsymbol{G}= \delta\boldsymbol{c}(\varphi)\boldsymbol{b}^T(\gamma)\in \mathbb{C}^{M\times I},
\end{align}where $\boldsymbol{G}=\boldsymbol{c}(\varphi)\boldsymbol{b}^T(\gamma)$

Let $y_m^{(n)}$ denote the signal received by the $m$-th antenna of the BS at time sample $n$, and define $\boldsymbol{y}^{(n)}=[y_1^{(n)},\cdots,y_M^{(n)}]^T$. In this paper, we assume that only LOS paths exist between the BS and the IRS as well as between the IRS and the users. Therefore, at time sample $n$, the signal received by the BS is given by
\begin{align}\label{eq:received signal spatial domain}
    \boldsymbol{y}^{(n)} & = \sum_{k=1}^{K}  \bar{\boldsymbol{G}}\text{diag}(\boldsymbol{\phi}^{(n)})\boldsymbol{h}_k \sqrt{p_k}s_{k}^{(n)} + \boldsymbol{z}^{(n)} \notag\\
    & = \delta \boldsymbol{G}\text{diag}({\boldsymbol{\phi}^{(n)}}) \sum_{k=1}^{K}   \beta_k\boldsymbol{b}(\theta_k) \sqrt{p_k}s_{k}^{(n)} + \boldsymbol{z}^{(n)},
\end{align}where $p_k$ denotes the transmit power of user $k$, $s_k^{(n)}\sim \mathcal{CN}(0,1)$ denotes the message of user $k$ transmitted at time sample $n$,  $\boldsymbol{\phi}^{(n)}=[\phi_{1}^{(n)},\cdots,\phi_{I}^{(n)}]^T$ with $\phi_i^{(n)}$ being the reflection coefficient of the $i$-th IRS element at time sample $n$, and $\boldsymbol{z}^{(n)} \sim \mathcal{CN}(\boldsymbol{0},\sigma^2\boldsymbol{I})$ denotes the BS noise at time sample $n$.

For simplicity of notation, define
\begin{align}
& \boldsymbol{\theta}=[\theta_1,\cdots,\theta_K]^T, \\ & \boldsymbol{B}(\boldsymbol{\theta})=[\boldsymbol{b}(\theta_1),\cdots,\boldsymbol{b}(\theta_K)], \\ & \hat{\boldsymbol{A}}^{(n)}(\boldsymbol{\theta})=\boldsymbol{G}\text{diag}({\boldsymbol{\phi}^{(n)}})\boldsymbol{B}(\boldsymbol{\theta}), \\ & \boldsymbol{x}^{(n)}=[\delta \beta_1\sqrt{p_1}s_{1}^{(n)},\cdots,\delta \beta_K\sqrt{p_K}s_{K}^{(n)}].\label{eqn:x}
\end{align}Then, the received signal at time sample $n$ in (\ref{eq:received signal spatial domain}) reduces to
\begin{align}\label{eq:received signal 1}
    \boldsymbol{y}^{(n)}  = \hat{\boldsymbol{A}}^{(n)}(\boldsymbol{\theta})\boldsymbol{x}^{(n)} + \boldsymbol{z}^{(n)}.
\end{align}

Note that we have ${\rm rank}(\hat{\boldsymbol{A}}^{(n)})=1$, $\forall n$, because ${\rm rank}(\boldsymbol{G})=1$ as can be observed from (\ref{eq:channel IRS BS}). Therefore, although (\ref{eq:received signal 1}) and (\ref{eq:received signal MIMO}) are in a similar form, the necessary condition (\ref{eqn:C2}) for MUSIC-based AOA estimation is not satisfied in our considered IRS-assisted systems. This indicates that it is impossible to apply the classic methods for estimating $\theta_k$'s based on the covariance matrix of the received signals given in (\ref{eq:received signal 1}). This is actually not surprising. Consider the ideal case when no noise exists at the BS, i.e., $\boldsymbol{z}^{(n)}=\boldsymbol{0}$, $\forall n$. According to (\ref{eq:received signal 1}), at time sample $n$, the signals received by antennas $m_1$ and $m_2$ have the following simple relation
\begin{align}
y_{m_2}^{(n)}=\frac{c_{m_2}(\varphi)}{c_{m_1}(\varphi)}y_{m_1}^{(n)}, ~ \forall m_2\neq m_1,
\end{align}where $c_m(\varphi)$ is the $m$-th element in the steering vector of the BS, i.e., $\boldsymbol{c}(\varphi)$. This implies that given the signal received by antenna $m_1$, the signal received by any other antenna $m_2\neq m_1$ provides no information about $\boldsymbol{\theta}$. In other words, in our considered IRS-assisted ISAC system, the phase differences among the signals received by different antennas at the BS only contain the information about the AOA of its incident signals from the IRS. However, we are interested in estimating the AOAs from the users to the IRS for localization. Although the spatial-domain multi-dimension signals do not work, we show a novel and interesting result in the rest of this paper: via properly creating virtual multi-dimension received signals of the BS in the temporal domain, we are able to estimate the AOAs of the LOS paths from users to the IRS based on the covariance matrix of these virtual signals, even if the received signals of only one BS antenna is utilized.

\section{A Temporal-Domain Approach for AOA Estimation}

In this section, we propose a novel approach for tackling the challenge shown in the previous section. Because signals received by different antennas of the BS provide no new information about $\boldsymbol{\theta}$, we propose to estimate $\boldsymbol{\theta}$ just based on the signals received by one antenna. For convenience, let us assume that the signals received by the first antenna of the BS are utilized. According to (\ref{eq:received signal 1}), at time sample $n$, the signal received by the first antenna of the BS is given as
\begin{align}\label{eq:received signal 2}
    y_1^{(n)}  = \boldsymbol{b}^T(\gamma)\text{diag}(\boldsymbol{\phi}^{(n)})\boldsymbol{B}(\boldsymbol{\theta})\boldsymbol{x}^{(n)} + z_1^{(n)},
\end{align}where $z_1^{(n)}\sim \mathcal{CN}(0,\sigma^2)$ denotes the noise at the first antenna of the BS at time sample $n$.

To estimate $\boldsymbol{\theta}$, we need to create a new received signal vector with dimension $L>K$ where the signals in the other $L-1$ dimensions all provide new information about $\boldsymbol{\theta}$ as compared to $y_1^{(n)}$. In the following, we provide an innovative method to create a virtual multi-dimension signal in the temporal domain for achieving the above goal.

Under our proposed scheme, every $L>K$ time samples form a block, and each user applies the repetition coding technique to transmit the same message within a block. Let $\tilde{s}_k^{(q)}$ denote the common message transmitted by user $k$ throughput the $q$-th block, $\forall q$. Then, we have \begin{align}
s_k^{((q-1)L+1)}=\cdots=s_k^{(qL)}=\tilde{s}_k^{(q)}, ~ \forall k.
\end{align}Let $Q$ denote the number of blocks that we will use for estimating $\boldsymbol{\theta}$. Then, the overall signal transmitted by user $k$ over $Q$ blocks (equivalently, $QL$ time samples), which is denoted as $\tilde{\boldsymbol{s}}_k=[s_k^{(1)},\cdots,s_k^{(QL)}]^T\in \mathbb{C}^{QL\times 1}$, is given as
\begin{align}\label{eq:repetition coding}
    \tilde{\boldsymbol{s}}_k \!=\! [\underbrace{\tilde{s}_{k}^{(1)},\!\cdots \!,\tilde{s}_{k}^{(1)}}_{\mathrm{Block}~1},\!\cdots \!,\underbrace{\tilde{s}_{k}^{(q)},\! \cdots\!,\tilde{s}_{k}^{(q)}}_{\mathrm{Block} ~q},\!\cdots\!,\underbrace{\tilde{s}_{k}^{(Q)},\!\cdots\!,\tilde{s}_{k}^{(Q)}}_{\mathrm{Block}~Q}]^T, \forall k.
\end{align}

According to (\ref{eq:received signal 2}), if the above repetition coding technique is used, the received signal of the first antenna at the $l$-th time sample in block $q$ is given as
\begin{align}\label{eq:received signal 3}
    y_1^{((q-1)L+l)}  = & \boldsymbol{b}^T(\gamma)\text{diag}(\boldsymbol{\phi}^{((q-1)L+l)})\boldsymbol{B}(\boldsymbol{\theta})\tilde{\boldsymbol{x}}^{(q)} \nonumber \\ & + z_1^{((q-1)L+l)}, ~ l=1,\cdots,L, ~ q=1,\cdots,Q,
\end{align}where $\tilde{\boldsymbol{x}}^{(q)}=[\delta \beta_1\sqrt{p_1}\tilde{s}_{1}^{(q)},\cdots,\delta \beta_K\sqrt{p_K}\tilde{s}_{K}^{(q)}]$ is obtained by substituting the user transmit messages in (\ref{eqn:x}) by (\ref{eq:repetition coding}) .

Define $\tilde{\boldsymbol{y}}_1^{(q)}=[y_1^{((q-1)L+1)},\cdots,y_1^{(qL)}]^T\in \mathbb{C}^{L\times 1}$ as the signal received by the first antenna throughout the $L$ time samples of the $q$-th block, $q=1,\cdots,Q$. It then follows that
\begin{align}\label{eq:received signal 4}
    \tilde{\boldsymbol{y}}_1^{(q)}  = \tilde{\boldsymbol{A}}^{(q)}(\boldsymbol{\theta})\tilde{\boldsymbol{x}}^{(q)} + \tilde{\boldsymbol{z}}^{(q)}, ~ q=1,\cdots,Q,
\end{align}where $\tilde{\boldsymbol{z}}^{(q)}=[z_1^{((q-1)L+1)},\cdots,z_1^{(qL)}]^T$, and
\begin{align}
\tilde{\boldsymbol{A}}^{(q)}(\boldsymbol{\theta})=\begin{bmatrix}
\boldsymbol{b}^T(\gamma)\text{diag}(\boldsymbol{\phi}^{((q-1)L+1)})\boldsymbol{B}(\boldsymbol{\theta})\\
\vdots  \\
\boldsymbol{b}^T(\gamma)\text{diag}(\boldsymbol{\phi}^{(qL)})\boldsymbol{B}(\boldsymbol{\theta})
\end{bmatrix}\in \mathbb{C}^{L\times K}.
\end{align}

For the temporal-domain multi-dimension received signals given in (\ref{eq:received signal 4}), we can properly design $\boldsymbol{\phi}^{((q-1)L+1)}\neq \boldsymbol{\phi}^{((q-1)L+2)}\neq \cdots \neq \boldsymbol{\phi}^{(qL)}$ to ensure that ${\rm rank}(\tilde{\boldsymbol{A}}^{(q)}(\boldsymbol{\theta}))=K$, $\forall q$. This is in sharp contrast to the spatial-domain multi-dimension received signals given in (\ref{eq:received signal 1}) where ${\rm rank}(\hat{\boldsymbol{A}}^{(n)})=1$. Now, the necessary conditions for the MUSIC algorithm given in (\ref{eqn:C1}) and (\ref{eqn:C2}) can be satisfied via setting $L>K$ and a property design of $\boldsymbol{\phi}^{((q-1)L+l)}$'s. The remaining issue is the necessary condition (\ref{eqn:C3}), because $\tilde{\boldsymbol{A}}^{(q)}(\boldsymbol{\theta})$ may change over blocks in (\ref{eq:received signal 4}). To satisfy condition (\ref{eqn:C3}), we let the IRS repeat the same reflecting pattern over different blocks, i.e.,
\begin{align}
\boldsymbol{\phi}^{((q-1)L+l)}=\bar{\boldsymbol{\phi}}^{(l)}, ~ l=1,\cdots,L, ~ q=1,\cdots,Q, \label{eqn:constant IRS}
\end{align}where $\bar{\boldsymbol{\phi}}^{(l)}$ is the constant IRS reflection coefficient pattern that is used in the $l$-th time sample of all the blocks. In this case, the temporal-domain received signals given (\ref{eq:received signal 4}) reduce to
\begin{align}\label{eq:received signal 5}
    \tilde{\boldsymbol{y}}_1^{(q)}  = \bar{\boldsymbol{A}}(\boldsymbol{\theta})\tilde{\boldsymbol{x}}^{(q)} + \tilde{\boldsymbol{z}}^{(q)}, ~ q=1,\cdots,Q,
\end{align}where
\begin{align}
\bar{\boldsymbol{A}}(\boldsymbol{\theta})=\begin{bmatrix}
\boldsymbol{b}^T(\gamma)\text{diag}(\bar{\boldsymbol{\phi}}^{(1)})\boldsymbol{B}(\boldsymbol{\theta})\\
\vdots  \\
\boldsymbol{b}^T(\gamma)\text{diag}(\bar{\boldsymbol{\phi}}^{(L)})\boldsymbol{B}(\boldsymbol{\theta})
\end{bmatrix} = [\bar{\boldsymbol{a}}(\theta_1),\cdots,\bar{\boldsymbol{a}}(\theta_K)],
\end{align}with
\begin{align}\label{eqn:steering vector IRS}
\bar{\boldsymbol{a}}(\theta_k)=\begin{bmatrix}\boldsymbol{b}^T(\gamma)\text{diag}(\bar{\boldsymbol{\phi}}^{(1)})\boldsymbol{b}(\theta_k) \\ \vdots \\ \boldsymbol{b}^T(\gamma)\text{diag}(\bar{\boldsymbol{\phi}}^{(L)})\boldsymbol{b}(\theta_k)\end{bmatrix}, ~ \forall k.
\end{align}

\begin{figure}
\centering
\includegraphics[scale=0.08]{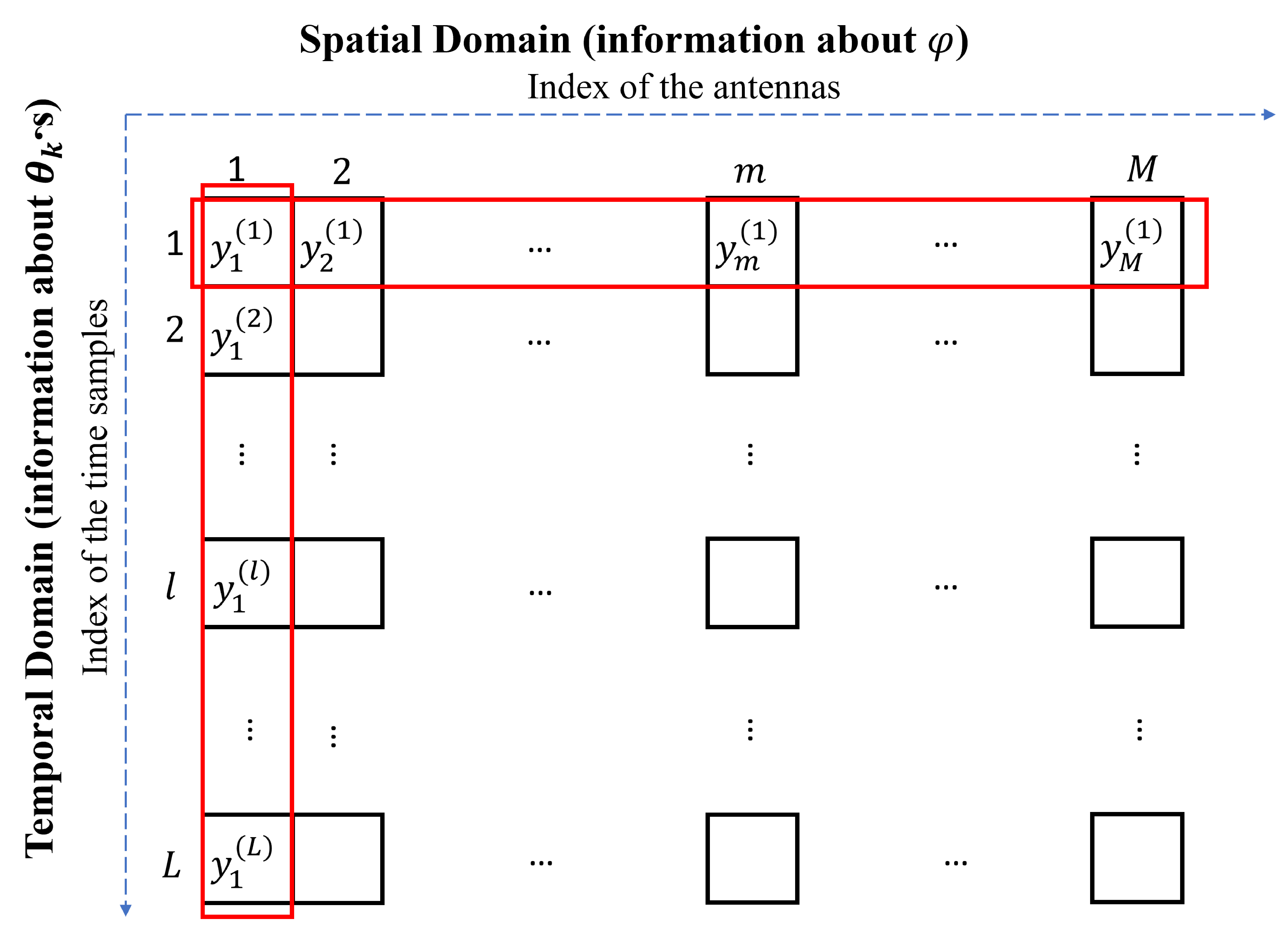}
\caption{Illustration of spatial-domain and temporal-domain signals: Spatial-domain signals provide information about $\varphi$, while our designed temporal-domain signals provide information about $\theta_1,\cdots,\theta_K$.}
\label{fig2}
\end{figure}

\begin{remark}
Mathematically, our designed temporal-domain received signals (\ref{eq:received signal 5}) describe a virtual system where $K$ users transmit messages to a BS with $L$ receive antennas, while the steering vector of the BS for a direction $\theta$ is given as $\bar{\boldsymbol{a}}(\theta)$. It can be shown that such a system satisfies the necessary conditions of the MUSIC algorithm given in (\ref{eqn:C1}), (\ref{eqn:C2}), and (\ref{eqn:C3}). The key to achieve the above goal is summarized as follows.
\begin{enumerate}
\item[1.] We select one antenna at the BS, e.g., antenna $1$, and accumulate its received signals in each block that consists of $L>K$ consecutive time samples to form a multi-dimension signal;
\item[2.] Each user adopts the repetition coding technique to transmit the same message over different time samples of the same block as shown in (\ref{eq:repetition coding});
\item[3.] The IRS adopts the same reflecting pattern over different blocks as shown in (\ref{eqn:constant IRS}), but different reflecting patterns over different time samples within each block, i.e., $\bar{\boldsymbol{\phi}}^{(1)}\neq \bar{\boldsymbol{\phi}}^{(2)}\neq \cdots \neq \bar{\boldsymbol{\phi}}^{(L)}$, to make ${\rm rank}(\bar{\boldsymbol{A}}(\boldsymbol{\theta}))=K$.
\end{enumerate}More interestingly, $\theta_1,\cdots,\theta_K$ in this virtual system are the AOAs that we are interested in under the IRS-assisted ISAC network. The relation between the spatial-domain signals and our designed temporal-domain signals is illustrated in Fig. \ref{fig2}.
\end{remark}

Because the necessary conditions given in (\ref{eqn:C1}), (\ref{eqn:C2}), and (\ref{eqn:C3}) are satisfied, we may apply the classic MUSIC algorithm for estimating $\boldsymbol{\theta}$ based on the received signals given in (\ref{eq:received signal 5}). The first step is to calculate the sample covariance matrix of the received signals as the estimation of the true covariance matrix. Specifically, the sample covariance matrix of the received signals obtained over $Q$ blocks is given as
\begin{align}\label{eqn:sample covariance matrix}
\boldsymbol{S}=\frac{1}{Q}\sum\limits_{q=1}^Q \left(\tilde{\boldsymbol{y}}_1^{(q)}\right)^H\tilde{\boldsymbol{y}}_1^{(q)}\in \mathbb{C}^{L\times L}.
\end{align}Then, define the eigenvalue decomposition (EVD) of $\boldsymbol{S}$ as $\boldsymbol{S}=\boldsymbol{U}\boldsymbol{\Lambda}\boldsymbol{U}^H$, where $\boldsymbol{\Lambda}={\rm diag}([\lambda_1,\cdots,\lambda_L]^T)$ whose diagonal elements are the eigenvalues of $\boldsymbol{S}$, and $\boldsymbol{U}=[\boldsymbol{u}_1,\cdots, \boldsymbol{u}_L]$ consists of the corresponding eigenvectors. Without loss of generality, assume that $\lambda_1\geq \lambda_2\geq \cdots \geq \lambda_L$. Then, define $\bar{\boldsymbol{U}}=[\boldsymbol{u}_{K+1},\cdots,\boldsymbol{u}_L]\in \mathbb{C}^{L\times (L-K)}$. Therefore, the spectrum used in the MUSIC algorithm is defined as
\begin{align}\label{eq:spectrum}
    P(\theta)=\frac{\bar{\boldsymbol{a}}(\theta)^H\bar{\boldsymbol{a}}(\theta)}{\bar{\boldsymbol{a}}(\theta)^H\bar{\boldsymbol{U}}\bar{\boldsymbol{U}}^H\bar{\boldsymbol{a}}(\theta)},\quad \forall \theta,
\end{align}where $\boldsymbol{a}(\theta)$ is given in (\ref{eqn:steering vector IRS}). At last, we can perform a one-dimension search to find the $K$ peaks of the above spectrum, and the corresponding angles will be the estimations of the AOAs of the LOS paths from the $K$ users to the IRS.

The overall method to estimate the AOAs of the signals from the users to the IRS is summarized in Table \ref{table1}.

\begin{table}[htp]
\begin{center}
\caption{Method for AOA Estimation in IRS-Assisted ISAC Network} \vspace{0.2cm}
 \hrule
\vspace{0.2cm}
\begin{itemize}
\item[1.] Divide $QL$ time samples into $Q$ blocks, while the $q$-th block consists of time samples $(q-1)L+1,\cdots,qL$, $q=1,\cdots,Q$.
\item[2.] Each user adopts the repetition coding technique to transmit the same message within each block, as shown in (\ref{eq:repetition coding}).
\item[3.] The IRS adopts the same reflection pattern over different blocks as shown in (\ref{eqn:constant IRS}), while different reflection patterns at different time samples within one block to make sure that ${\rm rank}(\bar{\boldsymbol{A}}(\boldsymbol{\theta}))=K$.
\item[4.] We accumulate the temporal-domain signals over each of the $Q$ blocks as shown in (\ref{eq:received signal 5}), and apply the MUSIC algorithm to estimate $\theta_1,\cdots,\theta_K$.
\end{itemize}
\vspace{0.2cm} \hrule \label{table1}
\end{center}
\end{table}

\section{Numerical Results}\label{sec:Numerical Results}
In this section, we provide numerical results to verify the effectiveness of our proposed MUSIC-based AOA estimation approach in IRS-assisted ISAC network. We assume that the coordinates of the BS and the IRS are $(0,0)$ and $(50,-50)$ in meters, respectively. Moreover, $K=3$ users are randomly located in a circle whose center is $(20,-20)$ in meters and radius is $30$ meters. Moreover, we assume that the IRS is equipped with $I=128$ reflecting elements. The steering vector of the IRS follows the ULA model as shown in (\ref{eqn:ULA1}), where the IRS reflecting element spacing is set as $\frac{\lambda}{2}$.

\begin{figure}
\centering
\includegraphics[scale=0.5]{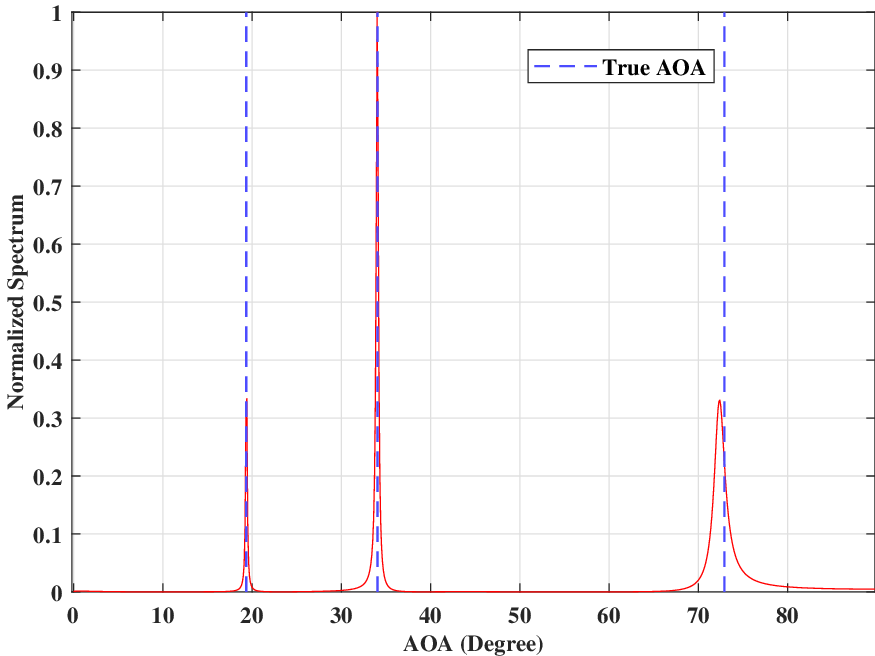}
\caption{Normalized spectrum of the MUSIC algorithm.}\vspace{-5pt}
\label{fig3}
\end{figure}

First, we show the normalized spectrum of the MUSIC algorithm in Fig. \ref{fig3}, which is defined as $P(\theta)/\max_\theta \left(P(\theta)\right)$ with $P(\theta)$ expressed as (\ref{eq:spectrum}). In this example, the true AOAs of the paths from the three users to the IRS are $\theta_1=72.9078^\circ$, $\theta_2=34.0409^\circ$, and $\theta_3=19.3314^\circ$, respectively. It is observed that the three AOAs leading to the peaks in the normalized spectrum are very close to the true AOAs.

\begin{figure}
\centering
\includegraphics[scale=0.5]{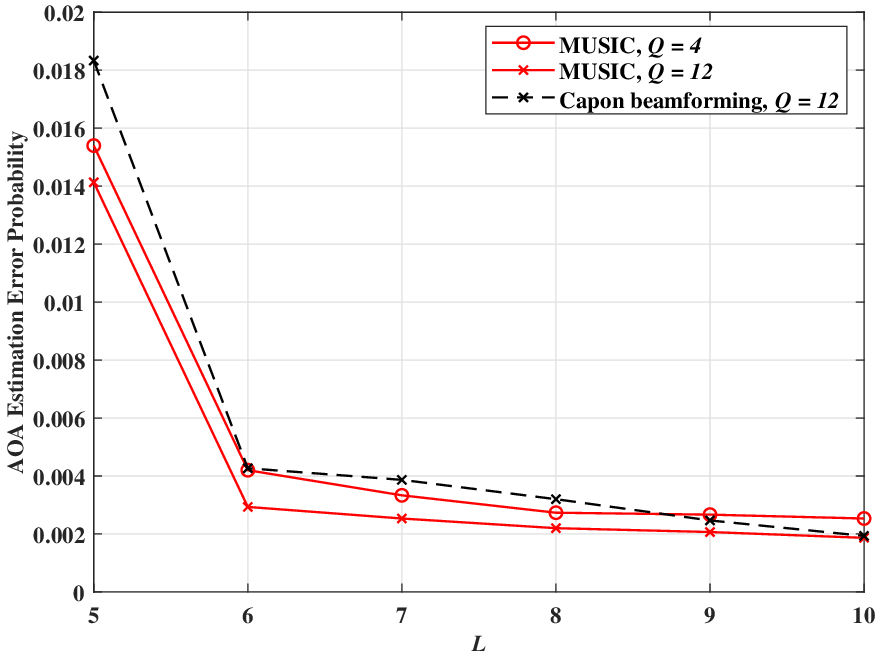}
\caption{Error probability performance of our proposed MUSIC-based method for AOA estimation in IRS-assisted ISAC network.}\vspace{-10pt}
\label{fig4}
\end{figure}

Next, we generate $5000$ realizations of the user locations, and record the number of error events in these realizations to calculate the error probability. Here, an error event is defined as the case that an estimated AOA is at least $1^\circ$ away from the true AOA. Fig. \ref{fig4} shows the AOA estimation error probability achieved by our proposed scheme with different values of $L$ and $Q$. We adopt the Capon beamforming method \cite{angle_survey,Capon} as the benchmark scheme. It is observed that under our proposed scheme, the error probability is around $0.4\%$ when $L=6$ and $Q=4$, i.e., $24$ time samples are used. Therefore, our proposed scheme is very powerful for AOA estimation. Moreover, under the benchmark scheme, when $Q=4$, the error probability is very large and not plotted in this figure. To achieve $0.4\%$ error probability, the benchmark scheme requires that $L=6$ and $Q=12$, i.e., $72$ time samples. Therefore, our scheme can estimate AOAs with much fewer time samples.

\section{Conclusions}
In this paper, we considered the estimation of the AOAs of the signals from the users to the IRS based on the signals received by the BS. Surprisingly, we showed that the MUSIC algorithm still works in this scenario. The key is to create the temporal-domain multi-dimension signals, where the signals across different dimensions provide useful information of the AOAs that we are interested in. Numerical results showed that our proposed scheme can estimate the AOAs in IRS-assisted ISAC network accurately.

\bibliographystyle{IEEEtran}
\bibliography{ISAC}
	
\end{document}